\documentclass[12pt]{article}
\usepackage{amsmath}
\usepackage{amssymb}
\newsymbol \blackbox 1004
\newcommand{\eh}{\hfill}\newlength{\sperr}

\newenvironment{proof}{{\settowidth{\sperr}{\bf\rm
Proof}%
\par\addvspace{0.3cm}\noindent\parbox[t]{1.3\sperr}
{\bf\rm P\eh r\eh o\eh o\eh f\eh }%
}}{\nopagebreak\mbox{} $\blackbox$\par\addvspace{0.3cm}}

\def\ze{\zeta}
\def\a{\alpha}
\def\b{\beta}

\def\G{\Gamma}

\def\clm{\mathcal M}

\def\s{\sigma}

\def\o{\omega}

\def\ov{\overline}

\def\vp{\varphi}
\def\ve{\varepsilon}
\def\z{\zeta}
\def\wh{\widehat}
\def\wt{\widetilde}
\def\ov{\overline}

\def\pr{\prime}
\def\BC{{\mathbb C}}
\def\BR{{\mathbb R}}
\def\BN{{\mathbb N}}

\def\cln{{\mathcal N}}

\newtheorem{Pa}{Paper}[section]
\newtheorem{Tm}[Pa]{{\bf Theorem}}
\newtheorem{La}[Pa]{{\bf Lemma}}
\newtheorem{Cy}[Pa]{{\bf Corollary}}
\newtheorem{Rk}[Pa]{{\bf Remark}}

\newtheorem{Ee}[Pa]{{\bf Example}}
\newtheorem{Dn}[Pa]{{\bf Definition}}
\newtheorem{Pn}[Pa]{{\bf Proposition}}
\newcommand{\CC}
{{\mathchoice {\setbox0=\hbox{$\displaystyle\rm C$}\hbox{\hbox
to0pt{\kern0.4\wd0\vrule height0.9\ht0\hss}\box0}}
{\setbox0=\hbox{$\textstyle\rm C$}\hbox{\hbox
to0pt{\kern0.4\wd0\vrule height0.9\ht0\hss}\box0}}
{\setbox0=\hbox{$\scriptstyle\rm C$}\hbox{\hbox
to0pt{\kern0.4\wd0\vrule height0.9\ht0\hss}\box0}}
{\setbox0=\hbox{$\scriptscriptstyle\rm C$}\hbox{\hbox
to0pt{\kern0.4\wd0\vrule height0.9\ht0\hss}\box0}}}}

\title{Weyl functions, inverse problem and special
solutions for the system auxiliary to the nonlinear optics
equation}
\author{Alexander Sakhnovich}
\date{}
\begin{document}
\maketitle

 {\bf Address}: Fakult\"at f\"ur Mathematik,
Universit\"at Wien, \\ Nordbergstrasse 15, A-1090 Wien, Austria \\
e-mail address: al$_-$sakhnov@yahoo.com\\

\begin{abstract}
A Borg-Marchenko type uniqueness theorem (in terms of the Weyl
function) is obtained here for the system auxiliary to the
$N$-wave equation. A procedure to solve inverse problem is used
for this purpose. The asymptotic   condition on the Weyl function,
under which the inverse problem is uniquely solvable, is completed
by the new and simple sufficient condition on the potential,
granting the fulfillment of this asymptotic condition. The
evolution of the Weyl function is discussed and the solution of an
initial-boundary value problem for the $N$-wave equation follows.
Explicit solutions of the system are obtained.
System with a shifted argument is treated. 
\end{abstract}

\section{Introduction} \label{intro}
\setcounter{equation}{0}
The well-known
integrable nonlinear optics ($N$--wave) equation has the form
\begin{equation} \label{0.3}
[D, u_t] - [\breve D, u_x] = \Big[[D, u], [\breve D
,u]\Big], \quad [D, u_t]:=Du_t-u_tD,
\end{equation}
where $D=D^*$ and $\breve D= \breve D^*$ are $m \times m$ diagonal matrices,
$u_x$ and  $u_t$ are the partial derivatives of the $m \times m$ matrix
function  $u(x,t)=u(x,t)^*$. Equation (\ref{0.3}) is the compatibility condition
of the auxiliary systems
\[
Y_x=G(x,t,z)Y, \quad Y_t=\breve G(x,t,z)Y, 
\]
where
\begin{equation} \label{i1}
G:=i z D - \zeta (x,t), \quad \breve G:=i z \breve D -\breve  \zeta (x,t), \quad \z=[D, u], 
\quad \breve \z=[\breve D, u].
\end{equation}
The case of three waves ($n=3$) interaction have been treated in the seminal paper
\cite{ZM}, using zero curvature representation. It was one of the first models 
to demonstrate the advantages of the  zero curvature representation. 
(See also \cite{AH} for the $N$-wave case.)
Several years later model  (\ref{0.3}) proved to be the first integrable system, which was studied
via  Riemann-Hilbert problem approach \cite{Sh}.
The $N$-wave equation is actively studied and used in optics, fluid dynamics and plasma physics.
It describes a variety of phenomena, including patterns and various instabilities,
and is also closely related to many other important nonlinear integrable wave equations.
For these and other  connections and applications see, for instance, the books and recent publications
\cite{ACh, AC, AS, Alb, CoG, DL, FT, Ge, Ka} and references therein.

We shall consider   the auxiliary to   (\ref{0.3})  system on the semiaxis:
\begin{equation} \label{0.1}
       Y_x(x,z)
=
   \bigl(
         i z D - \zeta (x)
   \bigr)
   Y(x,z),
\quad
     x \geq 0 \quad \Big(Y_x:=\frac{\, d Y}{dx} \Big),
\end{equation}
where, without loss of generality, we assume $D>0$:
\begin{equation}
      D = {\mathrm {diag}} \, \{ d_1^{\,}, d_2^{\,},
\ldots, d_m^{\,} \}, \quad
      d_1^{\,} > d_2^{\,} > \ldots > d_m^{\,} > 0;
\quad \zeta (x) = - \zeta^* (x).
    \label{0.2}
\end{equation}
Here  diag stands for diagonal matrix,
and $\zeta (x)$ is an $m \times m$ matrix function.
The study of system (\ref{0.1})  is basic for study and construction
of the solutions of the $N$-wave equation (for study of the difficult 
and important initial-boundary
value problem for  the $N$-wave equation, in particular).
Moreover, this system  is a natural generalization
of the well-known Dirac (called also Zakharov-Shabat or AKNS) system.
Various useful results and references on the scattering problem
for system (\ref{0.1}) one can find in \cite{AC, BC, BC1, BDT, BDZ, GeK, M}.
An extensive  amount of research
on the scattering problem for system (\ref{0.1})  with the
complex-valued entries $d_{k}$ of $D$ was done  by R. Beals and R.R.
Coifman.
The unique solvability of the problem on a
suitable dense set of scattering data was obtained, in particular.
The case (\ref{0.2}) of the positive $D$, that is considered here,
is less general, and therefore more explicit description of the
class of data for which the solution of the inverse problem exists
and a close to the classical procedure for solving this inverse problem
have been obtained in \cite{SaA2, SaA3, SaA4} in terms of the
generalized Weyl functions. Notice that various generalizations of
the Weyl functions are successfully used both in the inverse
scattering and in the inverse spectral  problems.

Here we develop further the results from \cite{SaA2, SaA3, SaA4}.
It is shown that the asymptotic conditions (\ref{1.40}) and (\ref{1.40'}) on the generalized 
Weyl function,
under which the inverse problem is uniquely solvable, are automatically fulfilled
for the integrable and two times differentiable $\ze$ with integrable derivatives. 
If conditions (\ref{1.40}) and (\ref{1.40'}) are valid for the Weyl function of the initial system,
they are valid also for systems with
a shifted argument, and they hold under the evolution of the Weyl function too.
In this way we obtain
sufficient conditions, when the inverse problem for system (\ref{0.1})
and an initial-boundary value problem in the quarterplane for
the $N$-wave equation (\ref{0.3}) are solvable, and our procedure for solving these problems
works.

A  recent series of papers by F. Gesztesy, B. Simon and
coauthors on the high energy asymptotics of the Weyl functions and local Borg-Marchenko type
uniqueness results has initiated a growing interest interest in this important domain 
(see \cite{CGR, GS, GZ, SaA5', Si, Si2} and references therein). The Weyl-Titchmarsh theory for
a non-self-adjoint case (the skew-self-adjoint  Dirac type system) has been studied
in \cite{CG, GKS, SaA1} and  the Borg-Marchenko type results for this system have been published in
\cite{SaA5''}. In this paper we obtain a Borg-Marchenko type theorem for another important
non-self-adjoint case, that is, system (\ref{0.1}).

Construction of the explicit solutions of the inverse problems and nonlinear equations
is of great interest and B\"acklund-Darboux transformation is  one of the most
fruitful methods to do it. The initial approach by B\"acklund and Darboux
have been greatly generalized and developed (see, for instance, \cite{Cie, D, Ges, GT, Gu, ZMi}
and references therein). We consider some applications and developments
of the version of the B\"acklund-Darboux transformation, which is called GBDT
(see \cite{SaA3', SaA5, SaA6}).

Section 1 is an introduction and Section 2 contains preliminaries
to make the paper self-sufficient. Section 3 contains
Theorem \ref{TmSolv}, which states that for the two times differentiable $\ze$ with integrable derivatives
asymptotic conditions (\ref{1.40}) and (\ref{1.40'}) on the Weyl functions are fulfilled.
It contains also a
Borg-Marchenko type Theorem \ref{BM}. Weyl functions for systems with a shifted argument are
treated in Section 4. Evolution of the Weyl function and solution of the initial-boundary
value problem are described in Remark \ref{Ev}. GBDT for system (\ref{0.1})
is discussed in Section 5. Lemma \ref{La3.4} is proved in Appendix A and  formula (\ref{4.7})
is proved in Appendix B.

The space $L^1(a, \, b)=L^1_{m \times m}(a, \, b)$ of $m \times m$
matrix functions on $(a, \, b)$ is equipped with the norm
$\|f\|_1=\int_a^b\|f(x)\|d x$, where the matrix norm is defined in
terms of the trace Tr by the equality $\|f(x)\|=\Big({\mathrm{Tr}}
\big(f(x)^*f(x)\big)\Big)^{\frac{1}{2}}$. We denote by $C^k(a,b)$
the class of $k$ times differentiable matrix functions.
\section{Preliminaries} \label{prel}
\setcounter{equation}{0} We shall consider system    (\ref{0.1}),
such that the inequalities
\begin{equation}
      \sup\limits_{0 < x < l} \,
          \bigl\|
                \zeta (x)
          \bigr\|
      < \infty
    \label{1.26}
\end{equation}
are true for each $l < \infty$. 
The $m \times m$ fundamental solution $w$ of
system (\ref{0.1}) is normalized by the condition
\begin{equation} \label{n.1}
w(0,z)=I_m, 
\end{equation}
where $I_m$ is the $m \times m$ identity matrix. A generalized Weyl function, later
called  a Weyl function, is introduced for  (\ref{0.1}) slightly
different from \cite{SaA4}.
\begin{Dn} \label{Weyl}
   A   Weyl function of system
(\ref{0.1}) is an $m \times m$ matrix function $\varphi$, such
that for some $M>0$ it is analytic in a lower semiplane $\Im z <
-M$, and the inequalities
\begin{equation}
      \sup\limits_{x \, \leq \, l, \ {\Im} \, z \, <
\, -M} \,
          \bigl\|
                w(x,z)
                \varphi (z)
                \exp \, (-i x z D )
          \bigr\|
    < \infty
   \label{1.4}
\end{equation}
hold for all $l < \infty$.
\end{Dn}
System (\ref{0.1}) with a bounded  on the semiaxis potential
$\zeta$:
\begin{equation}
      \sup\limits_{0 < x < \infty} \,
          \bigl\| \zeta (x) \bigr\|
    \leq M_0^{\,},
   \label{1.3}
\end{equation}
was treated in Lemma 1.1 \cite{SaA4}. For this case, it was proved
that a Weyl function always exists and admits normalization
\begin{equation}
         \varphi_{kj}^{\,} (z) \equiv 1  \, \mbox{
for   } \,  k = j, \quad
         \varphi_{kj}^{\,} (z) \equiv 0 \, \mbox{
for   } \,  k > j.
    \label{1.2}
\end{equation}
Moreover, by Theorems 1.1 and 2.1 \cite{SaA4} a normalized, as in
(\ref{1.2}), Weyl function of system (\ref{0.1}) with a bounded on
the semiaxis potential is unique. This Weyl function satisfies for
some $r>0$ the inequality
\begin{equation}
      \int\limits_0^\infty
                   \Bigl(
                        \exp \, (i  x \overline  z \,
D )
                   \Big)
                   \varphi (z)^*
                   w (x,z)^*
                   w (x,z)
                   \varphi (z)
                  \exp \,  \Big(
                        x(- i z D - r I_m^{\,} )
                        \Big)
      \, dx
    <  \infty,
    \label{1.1}
\end{equation}
where $\Im z<-M$. Inequality (\ref{1.1}) is somewhat similar to
the inequalities characteristic for the classical Weyl functions.
\begin{Dn} \label{DnInv}
The inverse spectral problem (ISpP) for system (\ref{0.1}), which
satisfies (\ref{1.26}),  is the problem to recover  the system,
i.e., to recover the matrix function $\z$, such that
\begin{equation}\label{R0}
\zeta (x) = - \zeta (x)^*, \quad \quad \zeta_{kk}(x) = 0,
\end{equation}
from a Weyl function.  We shall denote by $\Omega$ the operator
mapping the pair $D$ and $\varphi (z)$ into $\zeta$, that is, $\, \Omega
(D, \varphi) = \zeta$.
\end{Dn}
We no longer assume  that $\varphi (z)$ satisfies (\ref{1.2}).
\begin{Tm} \label{TmUniq}
\cite{SaA4}       For any matrix function $\varphi (z)$, which is
analytic and bounded in the semiplane $\Im z <-M$, and which has
the property
\begin{equation}
      \int\limits_{- \infty}^\infty
          \bigl(
               \varphi (z) - I_m^{\,}
          \bigr)^* \,
          \bigl(
               \varphi (z) - I_m^{\,}
          \bigr)
      \, d \lambda
      < \infty
\qquad
      (z = \lambda - i \eta, \quad \lambda \in \BR,
       \quad
       \eta > M),
    \label{1.27}
\end{equation}
there is at most one solution of the ISpP, i.e., $\Omega (D,
\varphi) $ is unique.
\end{Tm}
The existence of the  ISpP solutions was proved in Theorem 1.3
\cite{SaA4} under stricter conditions. Namely, we require that
\begin{equation}
            \sup \,
                 \Bigl\| \, z
                       \Big( \varphi (z) - I_m^{\,}
\Big)
                 \Bigr\|
            < \infty
         \quad
            (\Im  z < - M),
\label{1.40}
\end{equation}
and that for some matrix $\a$  for all lines $z = \lambda - i \eta$
with fixed  values $\eta > M$, we have
\begin{equation}
           z \left( \varphi (z) - I_m^{\,} - \alpha /
z \right)
            \in L_{m \times m}^2 (- \infty, \infty).
\label{1.40'}
\end{equation}
Without loss of generality we  suppose also that
\begin{equation}
      \det \, \varphi (z) \neq 0.
    \label{1.41}
\end{equation}
\begin{Tm} \cite{SaA4} \label{TmInv}
         Let the analytic matrix function
$\varphi$ satisfy (\ref{1.40})-(\ref{1.41}). Then $\vp$ is a Weyl
function of a unique system (\ref{0.1}), such that (\ref{R0})
holds.
\end{Tm}
Finally, let us give the procedure to recover $\zeta$ from $\vp$
(see the proof of Theorem 1.3 \cite{SaA4}). First notice that
relations (\ref{1.40})-(\ref{1.41}) yield
\begin{equation}
            \sup \,
                 \Bigl\| \, z
                       \Big( \varphi (z)^{-1} -
I_m^{\,} \Big)
                 \Bigr\|
            < \infty
         \quad
            (\Im  z < - M),
\label{1.43}
\end{equation}
and
\begin{equation}
           z \left( \varphi (z)^{-1} - I_m^{\,} +
\alpha / z \right)
            \in L_{m \times m}^2 (- \infty, \infty).
\label{1.43'}
\end{equation}
Therefore, we can introduce the $m \times m$ matrix function
\begin{equation}
       \Pi (x) = \frac{1}{\, 2 \pi i\,}
                 \int\limits_{- \infty}^\infty
                  \frac{1}{z}   \big(\exp \,
(ixzD)\big)         \,
                     \varphi (z)^{-1}
                 \,   d \lambda
\qquad
       (z = \lambda - i \eta,
        \quad
        \eta > M,
        \quad
        x \geq 0),
    \label{1.42}
\end{equation}
where the integral  is understood as the matrix function, which entries
are the norm limits in $L^2 (0,l)$ of the integrals from $a$ to $b$ ($a \to -\infty, \, b \to \infty$)
of the entries of $z^{-1} \big(\exp \,
(ixzD)\big)         \,
                     \varphi (z)^{-1}$.
 So $\Pi (x)\in L^2_{m \times m} (0,l)$ is defined on
each interval $(0,l)$. Recall that
\begin{equation}
      \frac{1}{\, 2 \pi i \,}
      \int\limits_{- \infty}^{\infty}
               \exp \, (ixzD)
           / z
      \, d \lambda
  \equiv I_m^{\,}
      \qquad
      (x \geq 0).
    \label{1.44}
\end{equation}
According to (\ref{1.43})--(\ref{1.44}), $\Pi (x)$ is twice
differentiable and the following properties hold:
\begin{equation}
            \Pi (0) = I_m^{\,}, \quad \Pi^\prime (0) =
- i D \alpha, \quad
               e^{- x M D } \, \Pi^\prime (x)
               \in  L_{m \times m}^2 (0, \infty)
               \quad (\Pi^\prime = \Pi_x),
    \label{1.45}
\end{equation}
\begin{equation}
\displaystyle               e^{-x M D } \, \Pi^{\prime\prime} (x)
               \in L_{m \times m}^2 (0, \infty).
    \label{1.45'}
\end{equation}
Now, let us introduce  the linear operator $S_l$, which is bounded
on $L^2_{m} (0,l)$:
\begin{equation}
      S_l^{\,} \, f = D^{-1} f + \int\limits^l_0
                                      s(x,u) f(u)
                                 \, du,
     \label{1.46}
\end{equation}
where $s(x,u) = \left\{ s_{kj}^{\,} (x,u) \right\}^m_{k,j=1}$; $0
\leq x, u \leq l$;
\renewcommand{\arraystretch}{1.5}
\[
      s_{kj}^{\,} (x,u) = \int\limits_\gamma
                              \theta_{kj}^{\,}
                              (v, u + d_k^{\,}
d_j^{-1} (v-x) )
                          \, dv
\] \begin{equation}
                     +
            \left\{
                 \begin{array}{lll}
                       d_k^{-1} \Pi^\prime_{kj}
(x-d_j^{\,} d_k^{-1} u)
                       & \mbox{ for }
                       & u \leq d_k^{\,} d_j^{-1} x,
\\
                       d_j^{-1}
                       \overline\Pi^{\,\prime}_{jk}
(u-d_k^{\,} d_j^{-1} x)
                       & \mbox{ for }
                       & d_k^{\,} d_j^{-1} x < u;
                 \end{array}
            \right.
     \label{1.47}
\end{equation}
$\theta(x,u) = \bigl\{
                      \theta_{kj}^{\,} (x,u)
               \bigr\}^m_{k,j=1}
             = \Pi^\prime (x)
               [ \, \Pi^\prime (u) \, ]^*
               D^{-1}$, \\
$\gamma$ is the interval $[\max \, (0, x-d_j^{\,} d_k^{-1} u) , \;
x]$.
      Sometimes we omit
"$l$" in $S_l^{\,}$ and write just $S$. The  operator $S$
satisfies the operator identity
\begin{equation} \label{1.48}
       A S - S A^* = i \, \Pi \, \Pi^*,
\end{equation}
where the operator $A$ acts in $L^2_m (0,l)$: $\, Af = i D
     \displaystyle{\int\limits_0^x}
                      f(u)
                   du$,
 and $\Pi$ acts from ${\BC }^m$ into $L^2_m
(0,l)$: $\, \Pi  g = \Pi (x)  g$. Using the operator identity, it
was shown in \cite{SaA4} that
\begin{equation}
       S_l^{\,} \geq \varepsilon (l) I
\qquad
       (\varepsilon > 0, \quad I - {\mathrm{identity}}
\, {\mathrm{operator}}).
    \label{1.49}
\end{equation}
By (\ref{1.45})--(\ref{1.49})  one gets the representation
\cite{GK}
\begin{equation}
      S^{-1} = V^* V,
\quad
      Vf = D^{1/2}f + \int\limits_0^x
                        V(x,u) f(u)  du,
    \label{1.51}
\end{equation}
where $ \int\limits_0^l
          \int\limits_0^x
              V (x,u)^*
              V(x,u)
          du
      dx
      < \infty$.
The operator $V^{-1}$ admits representation
\begin{equation}
       V^{-1}f
             = D^{-1/2}f + \int\limits_0^x
                              \Gamma(x,u)    f(u) du.
    \label{1.58}
\end{equation}
Moreover, according to formula (1.60) in \cite{SaA4} we get
\begin{equation}
      \begin{array}{rlll}
            V(u,x)^*    &=& -S_u^{-1} s(x,u) D^{1/2} &
\ (0<x \leq u),
        \\
            \Gamma (x,u) &=& s(x,u) D^{1/2}
                             +
\displaystyle{\int\limits_0^u}
                                   s(x,v) \,
                                   V(u,v)^*
                               \, dv
& \ (x \geq u>0),
      \end{array}
    \label{1.60}
\end{equation}
where $S_u^{-1}$ is applied to $s(x,u)$ columnwise. Formula
(\ref{1.60}) implies
\begin{equation}
      \Gamma (l,l) = D^{-1}
                     \bigl(
                          S_l^{-1}
                          s(x,l)
                     \bigr) (l) \,
                     D^{1/2}.
    \label{1.73}
\end{equation}
The potential $\zeta$ can be easily recovered from $\Gamma(l,l)$
\cite{SaA4}:
\begin{equation}
       \zeta (l)
    =  \bigl( \,
               \Gamma (l,l) \,
             - D \, \Gamma (l,l) D^{-1}
       \bigr)
       D^{1/2}.
    \label{1.72}
\end{equation}
Thus, formulas (\ref{1.42}),  (\ref{1.46}), (\ref{1.47}),
(\ref{1.73}), and (\ref{1.72}) determine $\zeta$. The proof of
(\ref{1.72})  in \cite{SaA4} is based on the representation of the
fundamental solution $w$:
\begin{equation}
    \label{1.75}
w(x,z)=D^{-\frac{1}{2}}\b(x)w_A(x,z),
\end{equation}
where $w_A$ is the transfer matrix function in the Lev Sakhnovich
form \cite{SaL1}-\cite{SaL3}:
\begin{equation}
    \label{1.76}
w_A(l,z)=I_m+iz\Pi^*S^{-1}\big(I-zA\big)^{-1}\Pi,
\end{equation}
and $\b(x)=\big(V\Pi\big)(x)$. By the last relation, in view of
(\ref{1.51}) and (\ref{1.60}), we have
\begin{equation}
    \label{1.77}
\b(l)=D^{\frac{1}{2}}\Big(\Pi(l)-\big(\Pi(u),S^{-1}s(u,l)\big)\Big),
\end{equation}
where $(\cdot , \cdot)$ denotes a matrix with the entries, which
are scalar products of the columns of matrices in the parenthesis,
that is
\begin{equation}
    \label{1.78}
\big(\Pi(u),S^{-1}s(u,l)\big)\Big)=\int_0^l\big(S^{-1}s(u,l)\big)^*\Pi(u)du.
\end{equation}
Now, consider system (\ref{0.1}) on the whole axis, and let $m
\times m$ matrix function $W$ satisfy (\ref{0.1}). Then,  matrix
function $\clm(x,z)=W(x,z)\exp(-ixzD)$ satisfies equation
\begin{equation} \label{1.78'}
       \clm_x(x,z)=
         i z\Big[ D,\clm(x,z)\Big]  - \zeta (x)\clm(x,z),
\quad
     -\infty<x <\infty,
\end{equation}
and vice versa. The function $\clm$ is defined by (\ref{1.78'}) up
to the right factor $\exp(ixzD)a(z)\exp(-ixzD)$. Normalization
conditions
\begin{equation}
    \label{1.79}
\lim_{x \to -\infty}\clm(x,z)=I_m, \quad \overline{\lim}_{x
\to\infty}\|\clm(x,z)\|<\infty
\end{equation}
are used to define $\clm$ uniquely. Sufficient conditions on the
potential $\ze$, under which relations (\ref{1.40}) and
(\ref{1.40'}) hold, follow from a particular case of the very
useful Theorem 6.1 \cite{BC}:
\begin{Tm} \label{TmBC} Suppose that the $m \times m$ potential $\ze$
is two times differentiable, i.e. $\ze(x) \in C^2(-\infty, \,
\infty)$, and that $\ze^{(k)}(x)=\frac{d^k\ze}{dx^k}\in
L^1(-\infty, \, \infty)$ for $k=0,1,2$. Then, for some $M>0$ the
analytic in $z$ matrix function $\clm(x,z)$, which satisfies
(\ref{1.78'}) and (\ref{1.79}), is well defined in the domain $\Im
z<-M$, the norm $\|\clm\|$ is uniformly bounded:
\begin{equation}
    \label{1.80}
\sup_{x \in (-\infty, \infty), \, \Im z<-M}\|\clm(x,z)\|<\infty,
\end{equation}
and uniformly with respect to $x$ we have
\begin{equation}
    \label{1.81}
\lim_{|z| \to \infty, \, \Im z<-M }\clm(x,z)=I_m.
\end{equation}
Finally, there is an $m\times m$ matrix function $\clm_1(x) \in
C^2(-\infty, \, \infty)$, such that
\begin{equation}
    \label{1.82}
\|\clm(x,z)-I_m-\frac{\clm_1(x)}{z}\|=O(z^{-2}) \quad (\Im z<-M),
\end{equation}
and that $\clm_1^{(k)}(x)\in L^1(-\infty, \, \infty)$ for $k=1,2$.
\end{Tm}
Scheme of the proof. First, the case $\|\ze \|_1<1$ is treated in
Theorem 3.8 \cite{BC}. The operator $K_{z,\, \ze}$ in the space of
bounded matrix functions $f(x)$ is introduced by the formula
\[
K_{z,\, \ze}f(x)=\int_x^\infty
\exp\big(i(x-y)zD\big)\big(\ze(y)f(y)\big)_+\exp\big(-i(x-y)zD\big)dy
\]
\begin{equation}
    \label{1.83}
-\int_{-\infty}^x
\exp\big(i(x-y)zD\big)\big(\ze(y)f(y)\big)_-\exp\big(-i(x-y)zD\big)dy,
\end{equation}
where $g_+$ ($g_-$) is the upper (lower) triangular part of the
matrix $g$, $g=g_+ +g_-$, and the main diagonal is included in
$g_-$.  One can see that $\|K_{z,\, \ze} \|\leq \|\ze \|_1$ and so
the operator $I-K_{z,\, \ze}$ is invertible. Moreover, it is
proved that $\clm(x,z)=(I-K_{z,\, \ze})^{-1}I_m$ satisfies
(\ref{1.78'}) and (\ref{1.79}), and  all the properties of $\clm$
follow. The general case $\|\ze \|_1<2^r$ ($r>0$) follows by
induction on $r$ (Theorem A \cite{BC}). In Theorem 6.1 matrix
functions $\clm_j(x)$ ($j=1,2$) are introduced by the equalities
$\lim_{x \to -\infty} \clm_j(x)=0$,
\begin{equation} \label{1.84}
\frac{d}{dx}\clm_{j-1}+\ze \clm_{j-1}=i[D,\clm_{j}^{od}], \quad
\clm_j^d(x)=-\int_{-\infty}^x \big(\ze(y)\clm_{j}^{od}(y)\big)^d
dy,
\end{equation}
where $\clm_0=I_m$, $\clm_j=\clm_j^{d}+\clm_j^{od}$, and the
entries of  $\clm_j$ and $\clm_j^{d}$ coincide on the main
diagonal, the entries of $\clm_j$ and $\clm_j^{od}$ coincide
outside the main diagonal. Then, relation (\ref{1.82}) follows
from equality (6.14) in  \cite{BC}:
\[
\clm(x,z)^{-1}\clm^{a}(x,z)=I_m- \int_x^\infty
\exp\big(i(x-y)zD\big)g(y)_+\exp\big(-i(x-y)zD\big)dy
\]
\begin{equation}
    \label{1.85}
+\int_{-\infty}^x
\exp\big(i(x-y)zD\big)g(y)_-\exp\big(-i(x-y)zD\big)dy,
\end{equation}
where $\clm^{a}(x,z)=\sum_{j=0}^2z^{-j}\clm_j(x)$ and
\[
g(x,z)=z^{-2}\clm^{-1}(x,z)\Big( \frac{d}{dx}\clm_2(x)+\ze(x)
\clm_2 (x)\Big).
\]
\section{Solvability of the inverse problem and
Borg-Marchenko type result} \label{Borg} \setcounter{equation}{0}
Theorem \ref{TmBC} yields our next theorem.
\begin{Tm} \label{TmSolv} Suppose that the $m \times m$ potential $\ze$
is two times differentiable, i.e. $\ze(x) \in C^2[0, \, \infty)$,
and that $\ze^{(k)}(x)=\frac{d^k\ze}{dx^k}\in L^1[0, \, \infty)$
for $k=0,1,2$. Then, for some $M>0$ a Weyl  function of system
(\ref{0.1}) satisfies conditions (\ref{1.40}) and (\ref{1.40'}).
\end{Tm}
\begin{proof}. Define $\ze(x)$ on the semi-axis $x<0$ so that
the conditions of Theorem \ref{TmBC} hold. Then we have
\begin{equation}
    \label{2.-2}
w(x,z)=\clm(x,z)\exp(ixzD)\clm(0,z)^{-1} \quad (x \geq 0).
\end{equation}
Hence, in view of Definition \ref{Weyl} and formulas (\ref{1.80})
and (\ref{2.-2}), the function $\vp(z)=\clm(0,z)$ is a Weyl
function. Now, it is immediate from (\ref{1.82}) that conditions
(\ref{1.40}) and (\ref{1.40'}) are fulfilled.
\end{proof}
From the procedure to solve ISpP,  a Borg-Marchenko type result
follows.
\begin{Tm} \label{BM}
         Let the analytic $m \times m$ matrix
functions $\varphi_1$ and  $\varphi_2$ satisfy
(\ref{1.40})-(\ref{1.41}). Suppose that on some ray $c \Im z =  \Re
z <0$ ($\, c \in \BR$, $\Im z <-M$) we have
\begin{equation} \label{2.1}
\varphi_1(z)^{-1}-\varphi_2(z)^{-1}=e^{-ilzD}O(z) \quad
{\mathrm{for}} \, |z| \to \infty .
\end{equation}
Then $\varphi_1$ and  $\varphi_2$ are Weyl functions of systems
(\ref{0.1}) with potentials $\zeta_1$ and $\zeta_2$, respectively,
which satisfy  (\ref{R0}) and the additional equality
\begin{equation} \label{2.2}
\zeta_1(x) \equiv \zeta_2(x) \quad (0<x<l).
\end{equation}
\end{Tm}
\begin{proof}.
The fact that $\varphi_1$ and  $\varphi_2$ are Weyl functions
follows from Theorem \ref{TmInv}. From (\ref{1.40})-(\ref{1.41})
follow relations (\ref{1.43}) and (\ref{1.43'}). According to the
classical results on the Fourier transform in the complex domain
(see, for instance, Theorem V in \cite{PW}), the function
$\displaystyle \frac{1}{z}\varphi(z)^{-1}$, where $\varphi(z)$
satisfies (\ref{1.43}) and (\ref{1.43'}), admits Fourier
representation. Moreover, taking into account also formula
(\ref{1.42}) and Plansherel's theorem, we get this representation,
for $z=\lambda -i \eta$ and fixed $\eta > M$, in terms of $\Pi$:
\begin{equation} \label{2.3}
\frac{1}{z}\varphi(z)^{-1}=i D
\int_0^{\infty}\big(\exp(-ixzD)\big) \Pi(x) dx,
\end{equation}
where $\big(\exp(-xMD)\big) \Pi(x) \in L_{m \times m}^2 (0,
\infty)$. As we have $\big(\exp(-xMD)\big) \Pi(x) \in L_{m \times
m}^2 (0, \infty)$, so equalities (\ref{2.3}) hold pointwise.
Hence, we can use (\ref{2.3}) to apply Phragmen-Lindel\"of
theorem. Namely, put
\begin{equation} \label{2.4}
F(z):=\big(\exp(ilzD)\big)\int_0^{l}\big(\exp(-ixzD)\big)\Big(
\Pi_1(x)-\Pi_2(x)\Big) dx,
\end{equation}
where $\Pi_1$ and $\Pi_2$ correspond via formula (\ref{1.42}) to
$\varphi_1$ and  $\varphi_2$, respectively. By (\ref{2.3}) and
(\ref{2.4}) we obtain
\[
F(z)=\frac{-i}{z}D^{-1}\big(\exp(ilzD)\big)\Big(\varphi_1(z)^{-1}-\varphi_2(z)^{-1}\Big)
\]
\begin{equation} \label{2.5}
-\int_l^{\infty}\Big( \exp\big(i(l-x)zD\big)\Big) \Big(
\Pi_1(x)-\Pi_2(x)\Big) dx,
\end{equation}
In view of (\ref{2.1}), the relation
\begin{equation} \label{2.6}
\left\|
\frac{-i}{z}D^{-1}\big(\exp(ilzD)\big)\Big(\varphi_1(z)^{-1}-\varphi_2(z)^{-1}\Big)
\right\|=O(1)
\end{equation}
is true. Recall that $z=\lambda -i \eta$,  $\eta > M$, and that \\
$\big(\exp(-xMD)\big) \Pi_k(x) \in L_{m \times m}^2 (0, \infty)$.
It follows that
\begin{equation} \label{2.7}
\left\| \int_l^{\infty}\Big( \exp\big(i(l-x)zD\big)\Big) \Big(
\Pi_1(x)-\Pi_2(x)\Big) dx \right\|=O(\frac{1}{\sqrt{\eta}}), \quad
\eta \to \infty .
\end{equation}
According to (\ref{2.5})-(\ref{2.7}), the matrix function $F(z)$
is bounded on the ray $c \Im z =  \Re z$ ($\Im z<-M$). It is immediate that
$F$ is bounded also on the axis $\Im z =-M$. Therefore, function
$F$ given by (\ref{2.4}) satisfies conditions of the
Phragmen-Lindel\"of theorem in the angles with the boundaries $\Im
z =-M$ and $c \Im z =  \Re z$ ($\Im z<-M$) in the lower semiplane. That is,
$F$ is bounded for $\Im z \leq -M$. It easily follows from
(\ref{2.4}) that $F$ is bounded for $\Im z > -M$ too, and that
$F(z) \to 0$ for $z=\ov z$ tending to infinity. So we derive $F
\equiv 0$. This implies that
\begin{equation} \label{2.8}
\Pi_1(x) \equiv \Pi_2(x) \quad (0<x<l).
\end{equation}
Finally, notice that by (\ref{1.46}),  (\ref{1.47}), and
(\ref{1.73}) the matrix function $\G(x,x)$ on the interval $[0, \,
l]$ is determined by $\Pi(x)$ on the same interval. Hence,
formulas (\ref{1.72}) and (\ref{2.8}) imply (\ref{2.2}).
\end{proof}
\section{System with a shifted argument} \label{shift}
\setcounter{equation}{0} In this section we shall consider system
(\ref{0.1}) with a shifted argument:
\begin{equation} \label{3.1}
          Y_x(x+ \s,z)
=
   \bigl(
         i z D - \zeta (x+\s)
   \bigr)
   Y(x+\s,z),
\quad
     x \geq 0 .
\end{equation}
Taking into account normalization condition (\ref{n.1}), for the
fundamental solution $w(x, \s, z)$ of (\ref{3.1}) we have
\begin{equation} \label{3.2}
w(x, \s, z) = w(x+\s,z)w(\s, z)^{-1}.
\end{equation}
By (\ref{3.2}) and Definition \ref{Weyl} the next proposition is
immediate.
\begin{Pn} \label{Pn3.1}
Let $\vp(z)$ be a Weyl function of system (\ref{0.1}). Then, the
matrix function
\begin{equation} \label{3.3}
\vp( \s, z) = w(\s, z)\vp(z)\exp(-i \s z D)
\end{equation}
is a Weyl function of system (\ref{3.1}).
\end{Pn}
Notice that according to (\ref{1.60}) the matrix function
$\G(x,u)$ does not depend on  the choice of $l$ $\,(l \geq x \geq
u>0)$ for  the domain of operators $S$ and $A$. Putting in
(\ref{1.60}) $x=u$, we get
\begin{equation}
            \Gamma (x,x) = s(x,x) D^{1/2}
                             +
\displaystyle{\int\limits_0^x}
                                   s(x,v) \,
                                   V(x,v)^*
                               \, dv .
    \label{3.4}
\end{equation}
By (\ref{1.47}), (\ref{1.49}), the first relation in (\ref{1.60})
and equality (\ref{3.4}), one can see that $\Gamma (x,x)$ is
continuous for $x>0$. (In fact, $\Gamma (x,x)$ is differentiable.)
Therefore, according to (\ref{1.72}) the matrix function $\zeta
(x)$ is continuous too. We shall put
\begin{equation} \label{3.4'}
\Gamma (0,0):=\lim_{x \to +0} \Gamma (x,x),
\end{equation}
and similar to (\ref{1.72})  assume
\begin{equation}
       \zeta (0)
    =  \bigl( \,
               \Gamma (0,0) \,
             - D \, \Gamma (0,0) D^{-1}
       \bigr)
       D^{1/2}.
    \label{1.72'}
\end{equation}
Now, we can express $\zeta(0)$ in terms of the matrix $\a$, which
is defined by $\vp$ via representation (\ref{1.43'}). For that
purpose introduce an $m \times m$ matrix $\wh \a=\{\wh
\a_{kj}\}_{k,j=1}^m$ via the entries $\a_{kj}$ of $\a$:
\begin{equation} \label{3.5}
\wh \a_{kj}:= \a_{kj} \quad {\mathrm{for}} \, k \leq j; \quad \wh
\a_{kj}:= -\ov \a_{jk} \quad {\mathrm{for}} \, k > j.
\end{equation}
\begin{Pn} \label{Pn3.2}
Let the analytic $m \times m$ matrix function $\varphi$ satisfy
(\ref{1.40})-(\ref{1.41}). Then, for $\zeta = \Omega(D, \vp)$ we have
\begin{equation}
       \zeta (0)
    = i \bigl( \,
     D \, \wh \a
     - \wh \a D \,  \bigr)
       .
    \label{3.6}
\end{equation}
\end{Pn}
\begin{proof}.
In view of (\ref{1.47}) and (\ref{3.4}) one obtains
\begin{equation}
\Big(\G(0,0)\Big)_{kj}=d_k^{-1}d_j^{\frac{1}{2}}\Pi^{\pr}_{kj}(0)
\, {\mathrm{for}} \, k \leq j; \quad
\Big(\G(0,0)\Big)_{kj}=d_j^{-1}d_j^{\frac{1}{2}}\ov
\Pi^{\pr}_{jk}(0) \, {\mathrm{for}} \, k > j.
    \label{3.7}
\end{equation}
Recall that according to (\ref{1.45}) we have
$\Pi^{\pr}(0)=-iD\a$, and so formulas (\ref{3.5}) and (\ref{3.7})
imply that $\G(0,0)=-i \wh \a D^{\frac{1}{2}}$. Hence, formula
(\ref{3.6}) follows from (\ref{1.72'}).
\end{proof}
\begin{Ee} \label{Eeconst}
Consider the simplest example  $m=2$ and
\begin{equation} \label{E1}
\zeta (x)\equiv \left[ \begin{array}{cc} 0 & -q \\ \ov q & 0
\end{array}
\right]= {\mathrm{const}},
\end{equation}
where const means a constant matrix. Calculate eigenvalues and
eigenvectors of $izD- \zeta $ to get
\begin{equation}
    \label{E2}
izD- \zeta =T(z)\Lambda(z) T(z)^{-1}, \quad
\Lambda(z)=\left[\begin{array}{cc} \lambda_1(z) & 0\\ 0 & \lambda_2(z)
\end{array}
\right],
\end{equation}
\begin{equation}
    \label{E2'}
T(z)= \left[\begin{array}{cc} 1 & (izd_2-\lambda_2(z))/ \ov q \\
(\lambda_1(z)-izd_1)/ q  & 1
\end{array}
\right],
\end{equation}
where $\lambda_k$ are the roots of equation
\begin{equation}
    \label{E3}
(izd_1-\lambda)(izd_2-\lambda)+|q|^2=0, \quad {\mathrm{i.e.}}
\end{equation}
\begin{equation}
    \label{E4}
\lambda_{1,2}=\frac{i}{2}\Big((d_1+d_2)z\pm\sqrt{(d_1-d_2)^2z^2+4|q|^2}\Big).
\end{equation}
In particular, we have
\begin{equation}
    \label{E5}
\lambda_k(z)-izd_k=(-1)^{k+1} 2i|q|^2\Big
(\sqrt{(d_1-d_2)^2z^2+4|q|^2}+ (d_1-d_2)z\Big)^{-1}.
\end{equation}
In view of (\ref{E2}) we get the fundamental solution
\begin{equation}
    \label{E6}
w(x,z)=T(z)\exp\{x\Lambda(z)\}T(z)^{-1}.
\end{equation}
From  (\ref{E5}) and (\ref{E6}) it follows that the matrix
function $\vp(z)=T(z)$ satisfies (\ref{1.4}), and so this $\vp$ proves a Weyl
function of system (\ref{0.1}) with $\zeta$ of the form
(\ref{E1}). Moreover, from (\ref{E2'}) and (\ref{E5}) it follows
that
\begin{equation}
    \label{E7}
\vp(z)=T(z)=I_2+\frac{i}{(d_1-d_2)z}\left[ \begin{array}{cc} 0 & q
\\ \ov q & 0
\end{array}
\right]+O\Big(\frac{1}{z^3}\Big).
\end{equation}
Therefore conditions (\ref{1.40})-(\ref{1.41}) are fulfilled.
\end{Ee}
\begin{Rk} \label{E}
Notice that Example \ref{Eeconst} describes the simplest case,
where conditions (\ref{1.40}) and (\ref{1.40'}) are fulfilled, but
conditions of Theorem \ref{TmSolv} are not valid.
\end{Rk}
Using Proposition \ref{Pn3.2} we obtain a similar result for a
system with a shifted argument.
\begin{Pn} \label{Pn3.3}
Let the analytic $m \times m$ matrix function $\varphi(z)$ satisfy
(\ref{1.40})-(\ref{1.41}), where
\begin{equation} \label{3.8}
\a^*=-\a .
\end{equation}
Then the matrix function $\varphi(\s,z)$ also admits
representation (\ref{1.40})-(\ref{1.41}), where $M(\s)=M+\ve$ for
an arbitrary fixed $\ve >0$, and where the matrix $\a(\s)$ is such
that
\begin{equation} \label{3.9}
\a(\s)^*=-\a(\s) .
\end{equation}
Moreover, for $\zeta = \Omega(D, \vp)$ we have
\begin{equation}
       \zeta (\s)
    = i \bigl( \,
     D \,  \a(\s)
     -  \a (\s)D \,  \bigr)
       .
    \label{3.10}
\end{equation}
\end{Pn}
\begin{Rk} \label{RkM}
If the conditions of Theorem \ref{TmSolv} are fulfilled, we have
\begin{equation} \label{r1}
\vp(\s, z)=\clm(\s,z), \quad \a(\s)=\clm_1(\s).
\end{equation}
According to the first relation in (\ref{1.84}) and equality
$\clm_0 \equiv I_m$, we get
\begin{equation} \label{r2}
[D,\clm_1(\s)]=-i\ze(\s)=i\ze(\s)^*, \quad
\clm_1^{od}(\s)^*=-\clm_1^{od}(\s).
\end{equation}
From the second equality in (\ref{1.84}) and  the first equality
in (\ref{r2}) it follows that the $k$-th diagonal entry of
$\clm_1$ has the form
\begin{equation} \label{r3}
\Big(\clm_1^{d}(\s)\Big)_{kk}=-i\int_{-\infty}^\s \sum_{j\in
\cln_k}(d_j-d_k)^{-1}
|\ze_{kj}(y)|^2dy=-\ov{\Big(\clm_1^{d}(\s)\Big)_{kk}},
\end{equation}
where $\cln_k=\{j\in \BN \, |\, 0<j \leq m, \, j\not=k \}$.
Formulas (\ref{r2}) and  (\ref{r3}) imply $\clm_1^*=-\clm_1$, and
so, using (\ref{r1}), we derive $\a^*=-\a$, $\a(\s)^*=-\a(\s)$.
Thus, under conditions of Theorem \ref{TmSolv}, equality
(\ref{3.8}) is true and the statement of Proposition \ref{Pn3.3}
follows from \cite{BC}. Still, the conditions of Proposition
\ref{Pn3.3}  are weaker than conditions of Theorem \ref{TmSolv}
(recall Example \ref{Eeconst}).
\end{Rk}
To prove  Proposition \ref{Pn3.3} we shall need some preparations.
In view of (\ref{1.46}) we can present $S^{-1}$ ($S=S_l$) in the
form
\begin{equation}
      S^{-1} f =Tf= Df + \int\limits^l_0
                                      T(x,u) f(u)du
                                 .
     \label{3.11}
\end{equation}
From $ST=I$, according to (\ref{1.46}) and (\ref{3.11}),  it
follows that
\begin{equation}
      s(x,u) D +D^{-1}T(x,u) + \int\limits^l_0 s(x,v)
                                      T(v,u) dv=0.
     \label{3.12}
\end{equation}
In particular, for the fixed values of  $u$  we shall assume
\begin{equation}
T(x,u)=-S^{-1}s(x,u)D,
                                 \label{3.13}
\end{equation}
and we shall define also $T(x,u)$ pointwise by the formula
\begin{equation}
T(x,u)=-D s(x,u)D+D \int\limits^l_0 \Big(S^{-1}s(v,x)\Big)^*
                                      s(v,u) dv \, D.
                                 \label{3.13'}
\end{equation}
Introduce now an $m \times m$ matrix function $K(x)$ by the
formula:
\begin{equation}
   K(x):= D^{-1}\Big(S^{-1}\Pi\Big)(x)=\Pi(x) +  D^{-1} \int\limits^l_0
                                      T(x,u) \Pi(u) du .
     \label{3.14}
\end{equation}
From the indentity  (\ref{1.48}), it follows that
$S^{-1}A-A^*S^{-1}=iS^{-1}\Pi\Pi^*S^{-1}$, i.e.,
\begin{equation}
   I_m +  D^{-1} \int\limits^l_u
                                      T(x,v)  dv+
\int\limits^l_x
                                      T(v,u)  dv   D^{-1}         =K(x)K(u)^*.
     \label{3.15}
\end{equation}
Using (\ref{3.13'}) and (\ref{3.14}), it is shown in the Appendix
1 that $K(x)$ is continuous and differentiable. The following
lemma is also proved in the Appendix 1.
\begin{La} \label{La3.4}
Let the analytic $m \times m$ matrix function $\varphi$ satisfy
(\ref{1.40})-(\ref{1.41}), and  (\ref{3.8}). Then, the relations
\begin{equation}
z\Big(w(l,z)\vp(z)
\exp(-ilzD)-I_m-\frac{i}{z}K(l)\Big(K^{\prime}(l)\Big)^*D^{-1}\Big)
\in L_{m \times m}^2 (- \infty, \infty),
    \label{3.16}
\end{equation}
and
\begin{equation}
\sup \left\|z\Big(w(l,z)\vp(z)
\exp(-ilzD)-I_m\Big)
\right\|<\infty
    \label{3.17}
\end{equation}
are true for $\Im z <-M-\ve$ (for any $\ve>0$).
\end{La}
\begin{proof} of Proposition \ref{Pn3.3}.
By formula (\ref{3.3}) and Lemma \ref{La3.4} we see that
$\varphi(l,z)$ satisfies (\ref{1.40})-(\ref{1.41}), and
\begin{equation}
\a(l)=iK(l)\Big(K^{\prime}(l)\Big)^*D^{-1}.
    \label{3.17'}
\end{equation}
Taking into account (\ref{3.13'}) we get, that $T(l,u)$ is
continuous in $u$ at $u=l$. Thus, putting $x=l$ and
differentiating both sides of (\ref{3.15}) with respect to $u$ at
$u=l$, we derive
\begin{equation}
K(l)\Big(K^{\prime}(l)\Big)^*=-D^{-1}T(l,l)D^{-1}.
    \label{3.18}
\end{equation}
Hence, according to (\ref{3.17'})and (\ref{3.18}), we have
\begin{equation}
\a(l)=-iD^{-1}T(l,l)D^{-1}.
    \label{3.19}
\end{equation}
From (\ref{1.45}), (\ref{1.47}), and (\ref{3.8}) it follows that
$s(l,l)=s(l,l)^*$. Therefore, formula (\ref{3.13'}) implies
$T(l,l)=T(l,l)^*$, and so, in view of (\ref{3.19}), we have
(\ref{3.9}). According to (\ref{3.9}), the equality $\wh \a(\s)
=\a(\s)$ is true, where
\[
\wh \a_{kj}(\s):= \a_{kj}(\s) \quad {\mathrm{for}} \, k \leq j;
\quad \wh \a_{kj}(\s):= -\ov \a_{jk}(\s) \quad {\mathrm{for}} \, k
> j.
\]
Now, as $\varphi(\s,z)$ satisfies (\ref{1.40})-(\ref{1.41}),
formula (\ref{3.10}) follows from Proposition \ref{Pn3.2}.
\end{proof}
\begin{Rk} \label{D}
Notice that under conditions of Theorem \ref{TmInv} formulas
(\ref{1.42}), (\ref{1.46}),  (\ref{1.47}),  (\ref{1.73}), and
(\ref{1.72}) define a solution $\z$ of the inverse problem even
without the requirement $d_k>d_j$ for $k>j$. That is, if $D>0$ and
$d_k \not= d_j$ for $k\not=j$, then $\z$, which is recovered from
$\vp$ by the mentioned above formulas, satisfies (\ref{R0}) and
defines such a system that (\ref{1.4}) holds. By Theorem
\ref{TmUniq} the solution of the inverse problem is unique. Recall
that we denote this solution $\z$ by $\Omega(D, \vp)$.
\end{Rk}
Consider now nonlinear optics ($N$-wave) equation (\ref{0.3}),
where $u=u^*$ and $D$ satisfies the second relation in
(\ref{0.2}).
\begin{Rk} \label{Ev} We assume for convenience that the entries of $u$ on the
main diagonal are identical zeros, i.e., $u_{kk}\equiv 0$.

Suppose first that the Weyl function $\vp$ is bounded and
satisfies (\ref{1.27}). Then \cite{SaA3}, for the case
\begin{equation} \label{wtD}
    \breve  D = {\mathrm {diag}} \, \{\breve  d_1^{\,}, \breve d_2^{\,},
\ldots, \breve d_m^{\,} \}, \quad
     \breve  d_1^{\,} > \breve d_2^{\,} > \ldots > \breve d_m^{\,} > 0,
\end{equation}
the initial condition
\begin{equation} \label{Du}
[D,u(x,0)]=\Omega(D,\vp)
\end{equation}
defines at most one continuously differentiable solution $u$ of
(\ref{0.3}) on the semi-band $x \geq 0, \quad  \o \geq t \geq 0$.

Consider now the more general case, where $D$ satisfies the second
relation in (\ref{0.2}), $\breve D>0$ and  $\breve d_k \not= \breve d_j$
for $k\not=j$, but the inequalities in (\ref{wtD}) do not necessarily hold. Define
the initial condition via Weyl function $\vp$ by formula
(\ref{Du}) and define the boundary condition via the same $\vp$:
\begin{equation} \label{wtDu}
[\breve D, u(0,t)]=\Omega(\breve D,\vp).
\end{equation}
Assume that  the analytic $m \times m$ matrix function $\varphi$
satisfies (\ref{1.40})-(\ref{1.41}), and  (\ref{3.8}). Then, we
have \cite{SaA3}: \\ a) The evolution of the Weyl function is
given by the formula
\begin{equation} \label{R1}
\vp(t,z)= R(t,z)\vp(z)\exp(-iz\breve D t),
\end{equation}
where
\begin{equation} \label{R2}
\frac{d R(t,z)}{d z}=\Big(iz \breve D- \Omega(\breve D, \vp) \Big)R(t,z),
\quad R(0,z)=I_m.
\end{equation}
Moreover,  the matrix functions $\vp(t,z)$  also satisfy
conditions (\ref{1.40})-(\ref{1.41}), and  (\ref{3.8}) for some
matrices $\a(t)$.\\ b) The matrix function $u(x,t)=u(x,t)^*$ is
well-defined by the relation
\begin{equation} \label{R3}
[D,u(x,t)]=\Omega\big(D,\vp(t,z)\big)
\end{equation}
and satisfies nonlinear optics equation (\ref{0.3}) and initial-boundary value
conditions (\ref{Du}) and (\ref{wtDu}).
\end{Rk}
\section{System with a shifted argument and Darboux
matrices} \label{Darb} \setcounter{equation}{0} From formula
(\ref{0.1}) it follows that $ \vp(\s,z)$ given by (\ref{3.3})
satisfies equation
\begin{equation} \label{4.1}
\frac{d \vp(\s,z)}{d \s}=iz\Big(D \vp(\s,z)-\vp(\s,z) D\Big)-
\zeta (\s) \vp(\s,z).
\end{equation}
The next proposition is also true.
\begin{Pn} \label{viceversa}
Let $\vp(\s, z)$ satisfy equation (\ref{4.1}), and let $\vp(0, z)$
be a Weyl function of system (\ref{0.1}). Then, the matrix
functions $\vp( \s, z)$ with fixed $\s>0$ are Weyl functions of
systems (\ref{3.1}).
\end{Pn}
\begin{proof}.
By (\ref{4.1})  one can see that $\vp(\s, z)e^{iz\s D}$  satisfies
(\ref{3.1}), i.e.,
\[
\vp(\s, z)e^{iz\s D}\vp(0, z)^{-1}=w(\s, z).
\]
Hence, $\vp(\s, z)$ has the form (\ref{3.3}), and our proposition
follows from Proposition \ref{Pn3.1}.
\end{proof}

Notice that equation (\ref{4.1}) coincides with the definition of
the Darboux matrix, which transforms solution $e^{iz x D}$ of the
auxiliary to nonlinear optics equation system with a trivial
potential $\zeta_0 =0$ into solution of  system (\ref{0.1}).
Therefore, by constructing Darboux matrices we obtain examples of
the Weyl functions. We propose below two schemes to construct
Darboux matrices. The first scheme is a particular case of the so
called GBDT (see \cite{SaA3', SaA5, SaA6} and references therein).
Namely, we shall introduce Darboux matrix as a transfer matrix
function (in Lev Sakhnovich form)  with additional dependence on
the variable $x$:
\begin{equation} \label{4.5}
w_A(x, z)=I_m-i\Pi(x)^*S(x)^{-1}(A-zI_m)^{-1}\Pi(x).
\end{equation}
Distinct from  formula (\ref{1.76}), both $\Pi(x)$ and $S(x)$ in
the GBDT method are differentiable matrix functions, where $\Pi$
is determined by the linear differential system. Correspondingly,
the factor $\Pi(x)^*$ above means multiplication by the matrix
adjoint to $\Pi(x)$. The second scheme is also constructed in the
spirit of the GBDT approach.
\\
{\bf Scheme 1.} This scheme is precisely GBDT for system
(\ref{0.1}) (see \cite{SaA2, SaA3'}). First, we fix an integer
$n>0$, two $n \times n $ parameter matrices $A$ and $S(0)$, and $n
\times m$ matrix $\Pi(0)$ such that
\begin{equation} \label{4.2}
AS(0)-S(0)A^*=i\Pi(0)\Pi(0)^*.
\end{equation}
Introduce $\Pi(x)$ and $S(x)$ for $x>0$ by the equations
\begin{equation} \label{4.3}
\Pi_x=-i A \Pi D + \Pi \zeta, \quad S_x=\Pi D \Pi^*,
\end{equation}
and put
\begin{equation} \label{4.4}
\wt \zeta (x)= \zeta (x)-\big(D
\Pi(x)^*S(x)^{-1}\Pi(x)-\Pi(x)^*S(x)^{-1}\Pi(x)D \big).
\end{equation}
Then, in the points of invertibility of $S(x)$, the transfer
matrix function $w_A$ satisfies \cite{SaA3'} the equation
\begin{equation} \label{4.7}
\frac{d}{dx}w_A(x, z)=\wt G(x,z) w_A(x, z)-w_A(x, z)G(x,z),
\end{equation}
where
\begin{equation} \label{4.8}
\wt G(x,z)=
         i z D -\wt \zeta (x), \quad
G(x,z)=         i z D - \zeta (x).
\end{equation}
By (\ref{4.2}) and (\ref{4.3}) we have also
\begin{equation} \label{4.14}
AS(x)-S(x)A^*=i\Pi(x)\Pi(x)^*.
\end{equation}
Further we assume that $S(0)>0$. Then, according to (\ref{4.3}),
we have $S(x)>0$, and so $S(x)$ is invertible. To make the paper
self-sufficient  we give the proof of formula (\ref{4.7}) in
Appendix 2.

{\bf Scheme 2.} Fix an interval $[a, \, b]$ and an $m
\times m$ weight matrix function $\rho (t)>0$, which is bounded on
this interval. Thus, the space $L^2_{m}(\rho)$ with the scalar
product $(f,g)=\int_a^b g(t)^*\rho(t) f(t) d t$ is generated. Let
$\zeta(x)$ be continuous, and define operators $A$ and $S(x)$ in
$L^2_{m}(\rho)$ by the formulas
\begin{equation} \label{4.9}
Af=tf(t), \quad S(x)f=cf+i \int_a^b\frac{w(x,t)^*w(x,y)}{t-y}
\rho(y)f(y)dy,
\end{equation}
respectively. Here we take the principal value of the integral in
(\ref{4.9}). Notice that $\frac{d}{dx}w(x,t)^*w(x,y)=-i(t-y)
w(x,t)^*Dw(x,y)$, i.e.,
\begin{equation} \label{4.10}
(t-y)^{-1}w(x,t)^*w(x,y)=(t-y)^{-1}-i\int_0^x w(u,t)^*Dw(u,y)du.
\end{equation}
Hence, we can rewrite the expression for $S$ from (\ref{4.9}) in
the form
\begin{equation} \label{4.11}
S(x)f=cf+i \int_a^b (t-y)^{-1} \rho(y)f(y)dy+\int_a^b \int_0^x
w(u,t)^*Dw(u,y)du \rho(y)f(y)dy.
\end{equation}
As $\rho$ is bounded and the operator $\int_a^b (t-y)^{-1} {\scriptstyle{\bullet}} \,
dy$ in $L^2_m$ is bounded, so the operator $\int_a^b (t-y)^{-1}
\rho(y) {\scriptstyle{\bullet}} \,dy$, and also operator $S(x)$, is bounded in
$L^2_{m}(\rho)$. Next, introduce operator $\Pi(x)$, acting from
$\BC^m$ into $L^2_{m}(\rho)$, and operator $\Pi(x)^*$:
\begin{equation} \label{4.12}
\Pi(x)f=w(x,t)^*f, \quad \Pi(x)^*g=\int_a^b w(x,y) \rho(y)g(y)dy.
\end{equation}
By (\ref{4.11}) and (\ref{4.12}) we have the second equality in
(\ref{4.3}), that is, $S_x=\Pi D \Pi^*\geq 0$. Now, choose $c$ so
that
\begin{equation} \label{4.13}
cI+i \int_a^b (t-y)^{-1} \rho(y)  {\scriptstyle{\bullet}} \, dy >0.
\end{equation}
As $S_x\geq 0$, by (\ref{4.13})  we have $S(x)>0$.  In view of
(\ref{0.1}),   definition (\ref{4.9}) of $A$ and definition
(\ref{4.12}), the first equality in  (\ref{4.3}) is true too.
Moreover, according to (\ref{4.9}) the identity (\ref{4.14})
holds. By (\ref{4.3}) and (\ref{4.14}) the equation (\ref{4.7}) is
satisfied, where $\wt G$ and $G$ are given via (\ref{4.4}) and
(\ref{4.8}), - see Appendix \ref{Ap2}. Using Definition
\ref{Weyl}, Proposition \ref{Pn3.1} and formula (\ref{4.7}) one
easily gets the next proposition.
\begin{Pn} \label{PnWeyl}
Let the Darboux matrix $w_A$ be defined via (\ref{4.5}), using
Scheme 1 or Scheme 2. Then the normalized fundamental solution of
the transformed system
\begin{equation} \label{4.14'}
\wt Y_x(x,z)=\wt G(x,z) \wt Y(x,z) = \bigl(
         i z D - \wt \zeta (x)
   \bigr)
\wt  Y(x,z)
\end{equation}
is given by the formula
\begin{equation} \label{4.15}
\wt w(x,z)=w_A(x,z)w(x,z)w_A(0,z)^{-1},
\end{equation}
where $w$ is the fundamental solution of the initial system
(\ref{0.1}). Suppose also that $\vp(z)$ is a Weyl function of
system (\ref{0.1}). Then the matrix function $w_A(0,z)\vp(z)$ is a
Weyl function of the system (\ref{4.14'}) and
\begin{equation} \label{4.16}
\wt \vp(\s, z)=w_A(\s, z)w(\s,z)\vp(z)\exp(-i \s z D)
\end{equation}
is a Weyl function of the system $\wt Y_x(x+\s,z)=\wt G(x+\s,z)
\wt Y(x+\s,z)$ with a shifted argument.
\end{Pn}
Next we shall consider 2 simple examples of Scheme 1, including
the case of $A$ non-diagonal.
\begin{Ee}
\label{A1} Let
\begin{equation} \label{4.20}
\ze(x)\equiv 0, \quad n=2, \quad \Pi(0)=\big[f_1 \quad f_2
\quad\ldots \quad f_m\big] \quad (f_k \in \BC^2).
\end{equation}
It follows from the first relation in (\ref{4.3}) that
\begin{equation} \label{4.21}
\Pi(x)=\Big[\exp(-ixd_1A)f_1 \quad \exp(-ixd_2A)f_2 \quad \ldots
\quad \exp(-ixd_mA)f_m\Big].
\end{equation}
It is immediate that
\begin{equation} \label{4.22}
\exp(-ixdA)={\mathrm{diag}}\{\exp(-ixda), \, \exp(-ixda)  \} \quad
{\mathrm{for}} \quad A= {\mathrm{diag}}\{a, \, a  \},
\end{equation}
and
\begin{equation} \label{4.23}
\exp(-ixdA)=\exp(-ixda)\left(I_2-ixd \left[
\begin{array}{cc}
0 & 1 \\ 0 & 0
\end{array}
\right] \right) \quad {\mathrm{for}} \quad A=\left[
\begin{array}{cc}
a & 1 \\ 0 & a
\end{array}
\right].
\end{equation}
Finally, assuming $a \not=\ov a$, formula (\ref{4.14}) implies
\[S=\{s_{kj}\}_{k,j=1}^2=i(a-\ov a)^{-1}\Pi\Pi^*\]
for the case (\ref{4.22}). For the case (\ref{4.23}) the same
formula implies
\[
s_{22}=i(a-\ov a)^{-1}\big(\Pi\Pi^*\big)_{22}, \quad s_{12}=\ov
{s_{21}}=(a-\ov a)^{-1}\Big(i\big(\Pi\Pi^*\big)_{12}-s_{22}\Big),
\]
\[
s_{11}=(a-\ov
a)^{-1}\Big(i\big(\Pi\Pi^*\big)_{11}+s_{12}-s_{21}\Big).
\]
Substitute explicit formulas for $\Pi$ and $S$, which are given
above, into (\ref{4.5}) and (\ref{4.4}) to obtain explicit
formulas for $w_A$ and $\wt \ze$.
\end{Ee}

Notice that $w_A(\s , z)$ admits representation
\begin{equation} \label{4.17}
w_A(\s ,z)=I_m+\frac{i}{z}\Pi(\s)^*S(\s)^{-1}\Pi(\s)
+O\Big(\frac{1}{z^2}\Big), \quad z\to \infty.
\end{equation}
From Proposition \ref{PnWeyl} and formula  (\ref{4.17}) follows
corollary.
\begin{Cy}
Let the conditions of  Proposition \ref{PnWeyl} be fulfilled, and
let a Weyl function $\vp$ of the initial system (\ref{0.1})
satisfy (\ref{1.40})-(\ref{1.41}) with the corresponding matrix
$\a$. Then, for some $M>0$, a Weyl function $\wt \vp(0, z)$ of the
transformed system (\ref{4.14'}) satisfies formulas
(\ref{1.40})-(\ref{1.41}), where  matrix $\a$ is substituted by
the matrix $\wt \a$:
\begin{equation} \label{4.18}
\wt \a=\a+i\Pi(0)^*S(0)^{-1}\Pi(0).
\end{equation}
\end{Cy}
From Propositions \ref{Pn3.3}  and \ref{PnWeyl} and from formula
(\ref{4.17}) we get.
\begin{Cy}
Let the conditions of  Proposition \ref{PnWeyl} be fulfilled, and
let a Weyl function $\vp$ of the initial system (\ref{0.1})
satisfy formulas (\ref{1.40})-(\ref{1.41}) and (\ref{3.8}) with
the corresponding matrix $\a$. Then, for some $M>0$, a Weyl
function $\wt \vp(\s, z)$ of the transformed system with a shifted
by $\s$ argument  satisfies formulas (\ref{1.40})-(\ref{1.41}),
(\ref{3.8}), where  matrix $\a(\s)$ is substituted by the matrix
$\wt \a(\s)$:
\begin{equation} \label{4.19}
\wt \a(\s)=\a(\s)+i\Pi(\s)^*S(\s)^{-1}\Pi(\s), \quad \wt \ze(\s)
=i(D \wt \a(\s)-\wt \a(\s)D).
\end{equation}
\end{Cy}
Formulas (\ref{4.19}) yield (\ref{4.4}).
\subsection*{Acknowledgements}
The work was supported by the Austrian Science Fund (FWF) under
Grant  no. Y330.
\begin{appendix}
\section{Appendix } \label{Ap1} \setcounter{equation}{0}
In this Appendix we shall obtain some properties of the matrix
function $K(x)$ defined in (\ref{3.14}), and using these
properties we shall prove Lemma \ref{La3.4}.
\begin{La}\label{LaA1}
Let the analytic $m \times m$ matrix function $\varphi$ satisfy
(\ref{1.40})-(\ref{1.41}), and  (\ref{3.8}). Then, $K(x)$ is twice
differentiable and satisfies equalities:
\begin{equation} \label{5.1}
I_m-\big(\Pi'(u),K(u)\big)=K(0)^*,
\end{equation}
\begin{equation} \label{5.2}
\big(D^{-1}\Pi{''}(u),K(u)\big)-\big(K'(0)\big)^*D^{-1}+ \Big(
I_m-\big(\Pi'(u),K(u)\big)\Big)D^{-1}\Pi'(0)=0.
\end{equation}
\end{La}
\begin{proof}.
According to (\ref{1.40})-(\ref{1.41}) and (\ref{1.42}), the
matrix function $\Pi(x)$ is two times differentiable. By
(\ref{3.13'}) the matrix function $T(x,u)+Ds(x,u)D$ is continuous
in both variables. Then, in view of formula (\ref{3.14}) $K(x)$ is
continuous. Consider now formula (\ref{3.15}). It follows that
$K(l)K(l)^*=I_m$. Hence, for the case $u=l$ formula (\ref{3.15})
yields
\begin{equation} \label{5.3}
K(x)=\left(I_m+\int_x^l T(v,u)dvD^{-1} \right)K(l).
\end{equation}
It follows from (\ref{5.3}) that $K(x)$ is differentiable and
\begin{equation} \label{5.4}
K'(x)=-T(x,l)D^{-1}K(l).
\end{equation}
Put now in (\ref{3.15}) $x=l$, multiply both sides from the right
by $\Pi'(u)$, and integrate the obtained expressions with respect
to $u$ from $0$ to $l$. We have
\begin{equation} \label{5.5}
\Pi(l)-\Pi(0)+D^{-1}\int_0^l\int_u^lT(l,v)dv\Pi'(u)du=K(l)\int_0^lK(u)^*\Pi'(u)du.
\end{equation}
Recall that by (\ref{1.45}) we have $\Pi(0)=I_m$ and change also the
order of integration in (\ref{5.5}). Then we derive
\begin{equation} \label{5.6}
\Pi(l)-I_m+D^{-1}\int_0^lT(l,v)(\Pi(v)-I_m)dv=K(l)\int_0^lK(u)^*\Pi'(u)du.
\end{equation}
Using (\ref{3.14}), rewrite (\ref{5.6}) in the form
\begin{equation} \label{5.7}
K(l)-I_m-D^{-1}\int_0^lT(l,v)dv=K(l)\int_0^lK(u)^*\Pi'(u)du.
\end{equation}
According to (\ref{3.15})  we get
\begin{equation} \label{5.8}
I_m+D^{-1}\int_0^lT(l,v)dv=K(l)K(0)^*.
\end{equation}
Recalling that $K(l)K(l)^*=I_m$ and using (\ref{5.7}) and
(\ref{5.8}), we finally obtain (\ref{5.1}). By (\ref{3.8}) and the
second relation in (\ref{1.45}) we have
\begin{equation} \label{5.9}
D^{-1}\Pi'(0)=\Big(D^{-1}\Pi'(0) \Big)^*.
\end{equation}
Hence, according to (\ref{1.47}),  $s(x,u)$ is continuous, and so
$T(x,u)$ is continuous. Moreover, in view of (\ref{3.13'}) and
(\ref{5.4}) one can see that $K$ is two times differentiable and
the entries of $K''(x)$ belong $L^2(0,l)$. Formula (\ref{1.47})
yields also
\begin{equation} \label{5.10}
D^{-1}\Pi'(x)=s(x,0).
\end{equation}
Put again in (\ref{3.15}) $x=l$, multiply both sides from the
right by $D^{-1}\Pi''(u)$, and integrate the obtained expressions
with respect to $u$ from $0$ to $l$:
\[
D^{-1}\big(  \Pi'(l)-\Pi'(0) \big)+D^{-1}\int_0^lT(l,v)D^{-1}\big(
\Pi'(v)-\Pi'(0) \big)dv
\] \[=
K(l)\Big( D^{-1}\Pi''(u), K(u) \Big).
\]
So, taking into account (\ref{5.10}),   we have
\[
D^{-1}\left(Ds(l,0)+\int_0^lT(l,v)s(v,0)dv
\right)-\left(I_m+\int_0^lT(l,v)dv \right)D^{-1}\Pi'(0)
\]
\begin{equation} \label{5.11}
=K(l)\Big( D^{-1}\Pi''(u), K(u) \Big).
\end{equation}
By  (\ref{3.11}), (\ref{3.13}), and (\ref{3.15}), we rewrite
(\ref{5.11}) as
\begin{equation} \label{5.12}
-D^{-1}T(l,0)D^{-1}-K(l)K^*(0)D^{-1}\Pi'(0)=K(l)\Big(
D^{-1}\Pi''(u), K(u) \Big).
\end{equation}
Recall that $S^{-1}=\big(S^{-1}\big)^*$  and that $T(x,u)$ is
continuous. Hence, it follows that $T(x,u)=T(u,x)^*$ and, in
particular, that $T(l,0)=T(0,l)^*$.  Therefore, using  (\ref{5.4})
and the equality $K(l)K(l)^*=I_m$, we obtain
\begin{equation} \label{5.13}
T(l,0)=-DK(l)\Big(K'(0)\Big)^*.
\end{equation}
From  (\ref{5.12}) and (\ref{5.13}) we get
\begin{equation} \label{5.14}
\Big(K'(0)\Big)^*D^{-1}-K^*(0)D^{-1}\Pi'(0)=\Big( D^{-1}\Pi''(u),
K(u) \Big).
\end{equation}
Finally, (\ref{5.1}) and (\ref{5.14}) imply (\ref{5.2}).
\end{proof}
\begin{proof}  of Lemma \ref{La3.4}.
For the proof of lemma we shall use representations (\ref{1.75})
and (\ref{1.76}). First consider expression $(I-zA)^{-1}\Pi$ from
(\ref{1.76}).  It is easy to see that
\begin{equation} \label{5.15}
\Big((I-zA)^{-1}\Pi\Big)(x)=\Pi(x)+izD\int_0^x\exp\big(i(x-u)zD\big)
\Pi(u)du.
\end{equation}
From  (\ref{2.3}) and (\ref{5.15}), using integration by parts, we
obtain
\[ \begin{array}{ccc}
\Big((I-zA)^{-1}\Pi\Big)(x) & &\\ =\Pi(x)+\big( \exp ixzD\big)
\Big(\vp(z)^{-1}&-&izD\int_x^{\infty} \exp\big(-iuzD\big)
\Pi(u)du\Big)
\end{array}
\]
\begin{equation} \label{5.16}
=\big( \exp ixzD\big) \vp(z)^{-1}+\frac{i}{z}D^{-1}\Big(\Pi'(x)+
\int_x^{\infty}\exp\big(i(x-u)zD\big) \Pi''(u)du \Big).
\end{equation}
By (\ref{1.76}) and (\ref{5.16}) we have
\[
w_A(l,z)\vp(z)\exp\big(-ilzD\big) =\vp(z)\exp\big(-ilzD\big) +iz
\Pi^*S^{-1} \Big(\exp\big(i(x-l)zD\big)
\]
\begin{equation} \label{5.17}
+\frac{i}{z}D^{-1}\Big(\Pi'(x)+
\int_x^{\infty}\exp\big(i(x-u)zD\big) \Pi''(u)du
\Big)\vp(z)\exp\big(-ilzD\big)\Big).
\end{equation}
To consider the asymptotics of the right-hand side of (\ref{5.17})
we take into account that $\Pi^*S^{-1}D^{-1}$ acts as the operator
$\int_0^l K(x)^* {\scriptstyle{\bullet}} \, dx$. Then, using integration by parts, we
get
\[
iz
\Pi^*S^{-1}\exp\big(i(x-l)zD\big)=K(l)^*-K(0)^*\exp\big(-ilzD\big)
\]
\begin{equation} \label{5.18}
-\int_0^lK'(x)^*\exp\big(ixzD\big)dx\exp\big(-ilzD\big).
\end{equation}
Use integration by parts again to rewrite (\ref{5.18}) in the form
\[
iz
\Pi^*S^{-1}\exp\big(i(x-l)zD\big)=K(l)^*+\frac{i}{z}K'(l)^*D^{-1}-K(0)^*\exp\big(-ilzD\big)
\]
\begin{equation} \label{5.19}
-\frac{i}{z}\Big(K'(0)^*D^{-1}\exp\big(-ilzD\big)+ q(l,z) \Big),
\end{equation}
where
\[
q(l,z):=\int_0^lK''(x)^*D^{-1}
\exp\big(ixzD\big)dx\exp\big(-ilzD\big),
\]
and so, for any $\ve>0$ we have
\begin{equation} \label{5.191}
\sup_{\Im \, z<-M-\ve}\|q(l,z)\|<\infty;
\end{equation}
and for the lines with the fixed values of $ \Im \, z$ we have
\begin{equation} \label{5.192}
q(l,z) \in L^2_{m\times m}(-\infty, \, \infty) \quad (\Im \,
z<-M-\ve, \quad  -\infty< \Re \,z < \infty) .\end{equation}
Consider now two other terms on  the right-hand side of
(\ref{5.17}) and take into account the second relation in
(\ref{1.45}) and (\ref{5.1}) as well as the asymptotics of $\vp$,
that is, formulas (\ref{1.40}) and (\ref{1.40'})  to obtain
\[
\vp(z)\exp\big(-ilzD\big)-\Pi^*S^{-1}D^{-1}\Pi'(x)\vp(z)\exp\big(-ilzD\big)
\]
\begin{equation} \label{5.193}
=K(0)^*\big(I_m
+\frac{i}{z}D^{-1}\Pi'(0)\big)\exp\big(-ilzD\big)+\frac{1}{z}q_1(l,z),
\end{equation}
where $q_1$ satisfies (\ref{5.191}) and (\ref{5.192}). Finally,
from integration by parts and asymptotics of $\vp$ it follows that
\[
- \Pi^*S^{-1}D^{-1}\int_x^{\infty}\exp\big(i(x-u)zD\big)
\Pi''(u)du \vp(z)\exp\big(-ilzD\big)
\]
\begin{equation} \label{5.194}
=\frac{i}{z}\Big(\int_0^lK(x)^*D^{-1}\Pi''(x)dx
+q_2(l,z)\Big)=\frac{i}{z}\Big(\Big(D^{-1}\Pi''(u), K(u)\Big)+
q_2(l,z)\Big),
\end{equation}
where $q_2$ satisfies (\ref{5.191}) and (\ref{5.192}). In view of
(\ref{5.1}) and (\ref{5.2}) the sum of the right-hand sides of
(\ref{5.19}), (\ref{5.193})  and (\ref{5.194}) equals
$K(l)^*+(i/z)K'(l)^*D^{-1}+q_3(l,z)/z$, where $q_3$ satisfies
(\ref{5.191}) and (\ref{5.192}). In other words we have
\begin{equation} \label{5.195}
w_A(l,z)\vp(z)\exp\big(-ilzD\big)
=K(l)^*+\frac{i}{z}K'(l)^*D^{-1}+\frac{1}{z}q_3(l,z).
\end{equation}
Further notice that in view of  (\ref{3.13}) and equality
$T(x,u)=T(u,x)^*$ we have
\begin{equation} \label{5.20}
\big(S^{-1}s(u,l)\big)^*=\big(-T(u,l)D^{-1}\big)^*=-D^{-1}T(l,u).
\end{equation}
According to  (\ref{1.77}) ,  (\ref{3.11}) and  (\ref{5.20})  it
follows that
\begin{equation} \label{5.21}
\b(l)=D^{\frac{1}{2}}D^{-1}\big(S^{-1}\Pi\big)(l).
\end{equation}
Compare (\ref{3.14}) and  (\ref{5.21}) to get
\begin{equation} \label{5.22}
\b(l)=D^{\frac{1}{2}}K(l).
\end{equation}
By (\ref{1.75}) ,  (\ref{5.195}), and  (\ref{5.21}) we obtain
(\ref{3.16}) and  (\ref{3.17}).
\end{proof}
\section{Appendix } \label{Ap2} \setcounter{equation}{0}
{\bf Proof} of formula (\ref{4.7}). \\ From (\ref{4.3}) and
(\ref{4.14}) it follows that
\begin{equation} \label{5.24}
\Big( \Pi^*S^{-1}\Big)_x = iD
\Pi^*A^*S^{-1}-\ze\Pi^*S^{-1}-\Pi^*S^{-1}\Pi D\Pi^*S^{-1}.
\end{equation}
By  (\ref{4.14}) we have $A^*S^{-1}=S^{-1}A-iS^{-1}\Pi
\Pi^*S^{-1}$, and so formula (\ref{5.24})  can be rewritten as
\[
\Big(\Pi^*S^{-1}\Big)_x =(D\Pi^*S^{-1}\Pi-\Pi^*S^{-1}\Pi
D-\ze)\Pi^*S^{-1}+iD \Pi^*S^{-1}A
\]
\begin{equation} \label{5.25}
=- \wt \ze \Pi^*S^{-1}+iD \Pi^*S^{-1}A,
\end{equation}
where $ \wt \ze$ is defined in (\ref{4.4}). Now, from from the
definition (\ref{4.5}) of $w_A$ and formulas (\ref{4.3}) and
(\ref{5.25}) it follows that
\[
\frac{dw_A}{dx}=-\wt \ze \big(w_A-I_m)+iD
\Pi^*S^{-1}A(A-zI_m)^{-1}\Pi-\big(w_A-I_m)(-\ze)
\]
\begin{equation} \label{5.26}
-
\Pi^*S^{-1}(A-zI_m)^{-1}A\Pi D.
\end{equation}
Substitute $A=(A-zI_m)+zI_m$ into the second and fourth terms on
the right-hand side to rewrite (\ref{5.26}) as
\begin{equation} \label{5.27}
\frac{dw_A}{dx}=(izD-\wt \ze)\big(w_A-I_m)-\big(w_A-I_m)(izD-
\ze)+D \Pi^*S^{-1}\Pi-\Pi^*S^{-1}\Pi D.
\end{equation}
Formulas (\ref{4.4}) and (\ref{5.27}) imply
\begin{equation} \label{5.28}
\frac{dw_A}{dx}=(izD-\wt \ze)\big(w_A-I_m)-\big(w_A-I_m)(izD-
\ze)+ \ze - \wt \ze,
\end{equation}
and (\ref{4.7}) is immediate.
\end{appendix}

\end{document}